\documentclass{article}
\usepackage{dcase2023_techrep,amsmath,graphicx,url,times,booktabs, tabularx}

\title{DIVIDED SPECTRO-TEMPORAL ATTENTION FOR SOUND EVENT LOCALIZATION AND DETECTION IN REAL SCENES FOR DCASE2023 CHALLENGE}

\name{Yusun Shul$^{1}$,
      Byeong-Yun Ko$^{2}$,
      Jung-Woo Choi$^{1}$ 
      }
\address{Korea Advanced Institute of Science and Technology, \\
        $^1$ School of Electrical Engineering, $^2$ Dept. of Mechanical Engineering, \\ 
        Yuseong-gu, Daejeon, Republic of Korea \\
         \{shulys, b.y.ko, jwoo\}@kaist.ac.kr\\  
 }

\begin{document}

\ninept
\maketitle

\begin{sloppy}

\begin{abstract}
Localizing sounds and detecting events in different room environments is a difficult task, mainly due to the wide range of reflections and reverberations. When training neural network models with sounds recorded in only a few room environments, there is a tendency for the models to become overly specialized to those specific environments, resulting in overfitting. To address this overfitting issue, we propose divided spectro-temporal attention. In comparison to the baseline method, which utilizes a convolutional recurrent neural network (CRNN) followed by a temporal multi-head self-attention layer (MHSA), we introduce a separate spectral attention layer that aggregates spectral features prior to the temporal MHSA. To achieve efficient spectral attention, we reduce the frequency pooling size in the convolutional encoder of the baseline to obtain a 3D tensor that incorporates information about frequency, time, and channel. As a result, we can implement spectral attention with channel embeddings, which is not possible in the baseline method dealing with only temporal context in the RNN and MHSA layers. We demonstrate that the proposed divided spectro-temporal attention significantly improves the performance of sound event detection and localization scores for real test data from the STARSS23 development dataset. Additionally, we show that various data augmentations, such as frameshift, time masking, channel swapping, and moderate mix-up, along with the use of external data, contribute to the overall improvement in SELD performance.
\end{abstract}


\section{Introduction}
\label{sec:intro}
The sound event localization and detection (SELD) task in DCASE challenge task 3 involves the classification of 13 different sound events and the estimation of their direction of arrival (DoA) \cite{Politis2022starss22}. Given the diverse nature of audio signals, characterized by a wide range of reflections and reverberations, addressing the issue of overfitting deep neural networks (DNNs) to specific room environments within the training dataset is of utmost importance. In previous years till 2021, the DCASE challenge task 3 has not permitted the utilization of external data. However, the use of external data and the incorporation of various types of augmentation emerged as the key contributions to the challenge of the DCASE2022 challenge. 

On the other hand, there is a growing trend of utilizing attention mechanisms separately across different dimensions, as evidenced in studies such as \cite{dividedatten}, \cite{Dasformer}, and \cite{deftan}. In the domain of video classification, \cite{dividedatten} extensively investigated different forms of attention and concluded that allocating attention to both temporal and spatial aspects independently yields valuable results. Similarly, for speech enhancement tasks, DASformer \cite{Dasformer} proposed the application of frame-wise spectral attention and band-wise temporal attention. Furthermore, DeFT-AN \cite{deftan} introduced a block composed of three distinct layers: a channel-related dense block, a frequency-related conformer, and a temporal conformer, which were tailored for speech enhancement.

Hence, we adopt the concept of divided attention to the spectro-temporal dimension in the SELD task and demonstrate its improvement over the baseline model that solely considers temporal context. We also utilize both the external data synthesized with AudioSet \cite{audioset} and augmentation methods such as frameshift \cite{freqshift}, time masking \cite{timemasking}, channel swap \cite{chswap}, and moderate mix-up \cite{M-mixup}, to improve the SELD performance along with the proposed DNN model.

% \textbf{Please note that headers for challenge and workshop submissions have different header.} Use the appropriate template!

\section{Proposed Method}
\label{sec:method}

\subsection{Implementation Details}
\label{ssec:imple}
Among the first-order Ambisonics (FoA) and tetrahedral microphone (MIC) audio formats available in STARSS23, the proposed model utilizes the FoA array signals. The audio signals having a sampling frequency of 24 kHz underwent a short-time Fourier transform with a hop length of 0.02s and a window size of 0.04s. Subsequently, the spectrograms were transformed into log-mel scales using 64 mel filter banks. Also, intensity vectors were calculated by multiplying the complex conjugation of the omnidirectional spectrograms, with the other directional array spectrograms, and taking the real part of the multiplication. These intensity vectors were then normalized using the total energy of the FoA signals. Therefore, the total number of input features was 7 channels consisting of four original spectrograms and three intensity components. Target DoA labels were converted into ACCDOA \cite{accdoa} and multi-ACCDOA \cite{multiaccdoa} formats. Experimental evaluations were carried out for both target formats. The model was trained for 500 epochs with a batch size of 128. The Adam optimizer was employed, using a learning rate of 0.001, $\beta_1=0.9$, and $\beta_2=0.999$. The input time length for the model was set to 5 seconds.

\begin{table}[ht]
\centering
\caption{Features of audio, preprocessing,  and input}
\begin{tabular}{l|l}
% \hline
% \multicolumn{2}{c}{Pre-processing}  \\ 
\hline
\multicolumn{2}{c}{Audio Features}\\
\hline
Audio Format & FoA (4 channels) \\
Sampling Frequency & 24kHz\\
\hline \hline
\multicolumn{2}{c}{Preprocessing Features}\\ \hline
Hop Size & 0.02s \\
Window Size & 0.04s \\
Number of Mel Filter Banks& 64\\
Target Label Resolution & 0.1s \\ %\cline{2-2}
\hline \hline
\multicolumn{2}{c}{Input Features}\\
\hline
Features & Log-mel, Intensity Vectors\\
Batch Size (B) & 128 \\
Input Feature Length (T) & 250\\
Input Channel (C) & 7 \\
Mel Frequency Bins (F) & 64\\
%\cline{2-2}
% Mel Filter Banks & 64 \\    
% Frame Length &   5s \\
% Window Length    & 0.04s  \\
% Hop Length       & 0.02s   \\ 
% Label Hop Length & 0.1s \\
\hline
\end{tabular}
\label{tb:preprocessing}
\end{table}

\subsection{External Data}
\label{ssec:extdata}
STARSS23 \cite{starss23_dataset} is an expanded version of STARSS22 \cite{starss22_dataset} additionally including real recordings captured in new room environments. While this additional data introduce more room diversity to the development training (dev-train) dataset, the development test (dev-test) data also have more diversity. Consequently, it remains challenging to prevent the overfitting of a model to the room impulse responses (RIRs) present in the training dataset. Therefore, it is still necessary to use an external dataset to mitigate overfitting on the training data. In this regard, AudioSet \cite{audioset} was utilized to synthesize external data through the convolution with RIRs in TAU-SRIR \cite{tau-srirDB} using the generator from \cite{data_gen}, in addition to the official synthetic dataset \cite{DCASE2022synth} based on FSD50k \cite{FSD50K}. The balanced dataset from AudioSet \cite{audioset} was utilized to ensure an equitable number of synthetic data across different classes. A total of 6 hours of the additional dataset with the maximum number of overlapping objects of two were synthesized and used for the training process along with the official synthetic data \cite{DCASE2022synth}.

\subsection{Data Augmentation}
\label{ssec:aug}

Since synthetic data often exhibit different data distributions compared to real recordings, it is beneficial to incorporate data augmentation along with external data. In light of this, we integrated various augmentations that have been proven to be effective for the SELD task, as demonstrated in \cite{M-mixup}. These augmentations include frameshift \cite{freqshift}, which involves shifting both the feature and label along the time axis using a randomly determined shifting parameter. Additionally, we employed time masking \cite{timemasking}, randomly masking features and DoA labels along the time axis. Furthermore, channel swap \cite{chswap} was utilized as a spatial augmentation method. Lastly, we incorporated moderate mix-up \cite{M-mixup}, an enhanced version of mix-up \cite{zhang2018mixup} specifically designed for the SELD task. This technique randomly selects a mixing ratio to blend two spectrograms and selects only one DoA with the larger mixing ratio as the new target label. All four augmentation methods were randomly selected and applied with random parameters generated for every batch and iteration.

\begin{figure}[t]
  \centering
  \centerline{\includegraphics[width=\columnwidth]{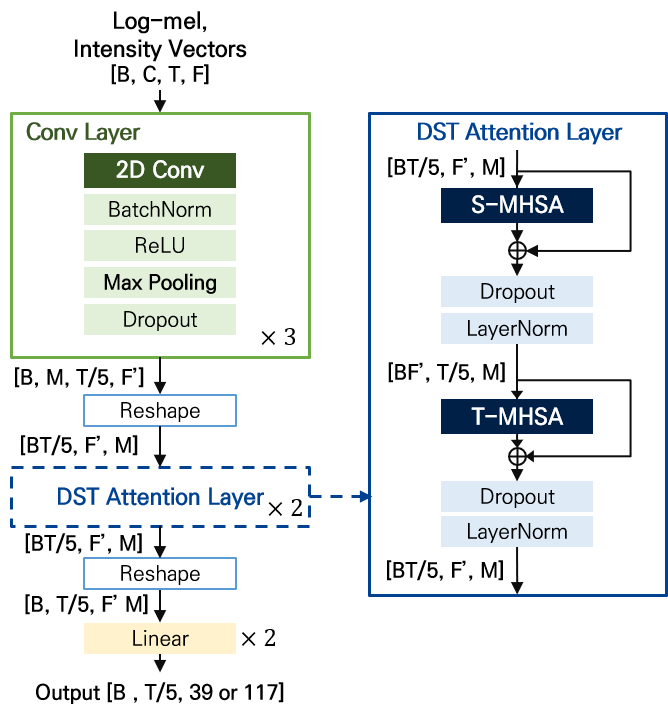}}
  \caption{Architecture of Divided Spetro-Temporal Attention Model}
  \label{fig:Archi_DividedST}
\end{figure}

\subsection{Divided Spectro-Temporal (DST) Attention}
\label{ssec:DSTatten}
The baseline model for DCASE2023 challenge task 3 \cite{baseline2023} incorporates several key components: convolutional blocks, bi-directional gated recurrent units (GRU), and multi-head self-attentions (MHSA). The convolutional blocks are responsible for amplifying the input channels from seven to 64, while also employing max pooling after each convolutional operation to reduce the spectral dimension to two. Additionally, to align the temporal information with the number of target labels, time pooling is utilized. The size of the time pooling kernel for the three convolutional layers is set to [$5,1,1$], respectively. The GRU and MHSA layers are specifically designed to capture and understand the temporal context from the spectral and channel embeddings. 

As a modification to enable the model to learn spectral information from the embeddings, we propose divided spectro-temporal attention for SELD, as depicted in Fig.~\ref{fig:Archi_DividedST}. This approach allows independent spectral attention from the channel embedding ($M=64$), separate from temporal attention. We reduced the kernel size of the frequency pooling layer to retain spectral information that can be used in the attention layer. The output from the convolution block of size (batch, channel, time, frequency)$=[B,~M,~T/5,~F^\prime]$ was then permuted for the divided spectro-temporal (DST) attention layer, such that the temporal dimension is merged to the batch dimension ($[B T/5,~F^\prime,~M$]). Then, spectral-MHSA (S-MHSA) was applied through attention maps of size ($F^\prime \times F^\prime $). The output from the S-MHSA layer was then merged with the input via the skip connection. The merged signal was then processed by the dropout layer and layer normalization. For the temporal-MHSA (T-MHSA), the frequency and time dimensions were swapped to yield the tensor size of $[B F^\prime,~T/5,~M]$. T-MHSA was then applied using attention maps of size $(T/5\times T/5)$, followed by the addition of attention input, dropout, and layer normalization. %Each input of MHSA layers was added up with residuals as in \cite{dividedatten}. 
The number of heads for MHSA was eight, and DST attention layers were repeated twice. The output of the final DST attention layer was reshaped to have the size $[B, T/5, F^\prime M]$ and fed into the two linear layers. The linear layers then reduced the size of embedding from $F^\prime M$ to that of the target embedding, which is 39 for ACCDOA and 117 for multi-ACCDOA.
The size of the spectral attention affects the SELD performance as presented in Table~\ref{tab:DSTwSdims}. The size of spectral dimension was varied ($F^\prime=8, 16, 32, 64$) by reducing the frequency pooling size as presented in Table~\ref{tab:train_param}. The final model was determined to have $F^\prime=16$.
% 
% \textbf{ADD equations of S-MHSA and T-MHSA}
\begin{table}[t!]
\centering
\caption{Training Parameters. In frequency pooling, $[c_1,~c_2,~c_3]$ represents pooling sizes in three convolution layers, respectively.}
\begin{tabular}{c|c|c}
\hline
\textbf{Model} & \textbf{Frequency Pooling} & \textbf{Spectral Dim.} (F')\\ \hline \hline
Baseline & [4,2,2] & 2\\ \hline
DST & [4,2,1] & 8\\
Attention & [2,2,1] & 16\\
& [2,1,1] & 32 \\
& [1,1,1] & 64\\
% \hline \hline
% \multicolumn{2}{c|}{Chnnel Embedding (M)} & 64\\
\hline
\end{tabular}
\label{tab:train_param}
\end{table}

\section{Results}
\label{sec:results}
The investigation results on DST attention, augmentation methods, and utilization of external data are summarized in Table~\ref{tab: DST_attention_results}. The results demonstrate that incorporating DST attention significantly enhances the SELD performance compared to the baseline model for both the single-ACCDOA and multi-ACCDOA tasks. Moreover, the performance consistently improves when augmentation techniques are applied, and these gains are further amplified when external data is employed.

Additionally, as discussed in Section \ref{ssec:DSTatten}, the size of spectral attention has an impact on SELD performance (Table~\ref{tab:DSTwSdims}). When the spectral dimension size is increased, the SELD score is improved for all cases of multi-ACCDOA training and up to $F^\prime=16$ for single-ACCDOA training, compared to the baseline shown in Table~\ref{tab: DST_attention_results}. However, increasing the spectral dimension beyond $F^\prime=16$ for the single-ACCDOA yields negative effects. It is worth noting that excessive spectral dimension ($F^\prime>16$) reduces the performance.

In Table~\ref{tab:DSTwSdims}, we can see that utilizing multi-ACCDOA outperforms single-ACCDOA when no augmentation or external data are involved with the training. However, with augmentation and external data, employing single-ACCDOA yields better outcomes, as shown in Table~\ref{tab: DST_attention_results}. Consequently, the final submitted version is the model trained using single-ACCDOA, four types of augmentation techniques, six hours of additional external data, and a spectral dimension size of 16 for the DST attention model. 

The SELD performances of the submitted models on the dev-test data are displayed in Table~\ref{tab:Submit_res}. Submission 2 is the model trained up to epoch 500, while submission 1 is a refinement of submission 2 through the training up to 700 epochs. On the other hand, submission 3 is the same model as submission 2 but is trained with different random initializations. It is noteworthy that all three submissions surpass the baseline model in terms of both event detection and localization, showing improved performances in all four metrics.

\begin{table}[hbt]
\caption{Experimental results of the proposed DST attention model on dev-test data in comparison with the baseline model.}
\label{tab: DST_attention_results}
\resizebox{\columnwidth}{!}{%
\setlength\tabcolsep{2.5pt}
\begin{tabular}{cccc|ccccc} 
\hline
\textbf{Model} & \textbf{\begin{tabular}[c]{@{}l@{}}ACC- \\ DOA\end{tabular}} & \textbf{Aug.} & \textbf{\begin{tabular}[c]{@{}l@{}}Ext. \\ Data\end{tabular}} & \multicolumn{1}{l}{\textbf{\begin{tabular}[c]{@{}l@{}}SELD \\ Score\end{tabular}}} & \multicolumn{1}{l}{\textbf{ER}} & \multicolumn{1}{l}{\textbf{F-score}} & \multicolumn{1}{l}{\textbf{LE}} & \multicolumn{1}{l}{\textbf{LR}} \\ \hline
Baseline & Multi & - & - & 0.4791 & 0.570 & 29.90 & 22.00 & 47.70 \\
& Single & - & - & 0.4642 & 0.615 & 33.62 & 22.88 & 54.91 \\ \hline
DST & Multi & - & - & 0.4345 & 0.580 & 39.50 & 20.03 & \textbf{55.83} \\
Attention &  & O & - & 0.4215 & 0.525 & 40.45 & 17.31 & 53.06 \\
($\mathbin{F^\prime=16}$) &  & O & O & 0.4329 & 0.555 & \underline{41.48} & \underline{16.92} & 50.26 \\
\cline{2-9}
& Single & - & - & 0.4423 & 0.570 & 37.16 & 20.30 & 54.20 \\
& & O & - &\underline{0.4209} &\textbf{0.505} & 39.87 & 17.95 & 52.24 \\
& & O & O & \textbf{0.4078} &\underline{0.510} &\textbf{41.96} & \textbf{16.54} &\underline{55.10}  \\ \hline
\end{tabular}
}
\end{table}

% \begin{table}[bt]
% \caption{Experimental results of the proposed DST attention on dev-test data in comparison with the baseline model}
% \label{tab: DST_attention_results}
% \resizebox{\columnwidth}{!}{%
% \begin{tabular}{cccc|cccc} %# {}안에 정렬 및 구분자 지정 
% \hline
% \textbf{Model} & \textbf{ACCDOA} & \textbf{Aug.} & \textbf{\begin{tabular}[c]{@{}c@{}}Ext. \\ Data\end{tabular}} & \multicolumn{1}{c}{\textbf{ER}} & \multicolumn{1}{c}{\textbf{F-score}} & \multicolumn{1}{c}{\textbf{LE}} & \multicolumn{1}{c}{\textbf{LR}} \\ \hline
% Baseline & Multi & - & -  & 0.570 & 29.90 & 22.00 & 47.70 \\
% & Single & - & - & 0.615 & 33.62 & 22.88 & 54.91 \\ \hline
% DST Attention & Multi & - & - & 0.580 & 39.50 & 20.03 & \textbf{55.83} \\
% ($F^\prime=16$)&  & O & - & 0.525 & 40.45 & 17.31 & 53.06 \\
% &  & O & O & 0.555 & \underline{41.48} & \underline{16.92} & 50.26 \\
% \cline{2-8}
% & Single & - & -  & 0.570 & 37.16 & 20.30 & 54.20 \\
% & & O & - &\textbf{0.505} & 39.87 & 17.95 & 52.24 \\
% & & O & O &\underline{0.510} &\textbf{41.96} & \textbf{16.54} &\underline{55.10} \\ \hline
% \end{tabular}
% }
% \end{table}

\begin{table}[bt]
\centering
\caption{SELD performances of DST attention with various $F^\prime$ on dev-test data (augmentation and external data were not used)}
\label{tab:DSTwSdims}
\resizebox{\columnwidth}{!}{%
\begin{tabular}{ccccccc}
\hline
\textbf{ACCDOA} & \textbf{$F^\prime$}  & \multicolumn{1}{l}{\textbf{\begin{tabular}[c]{@{}l@{}}SELD\\ Score\end{tabular}}} & \multicolumn{1}{c}{\textbf{ER}} & \multicolumn{1}{c}{\textbf{F-score}} & \multicolumn{1}{c}{\textbf{LE}} & \multicolumn{1}{c}{\textbf{LR}} \\
 \hline
Multi & 8 & 0.4422 & 0.595 & 37.42& 19.48 & 56.01 \\
& 16 & \textbf{0.4345} & 0.580 & \textbf{39.50} & 20.03 & \textbf{55.83} \\
& 32 & 0.4517 &\textbf{0.570} & 35.86 & 18.24 & 50.60 \\
& 64 & 0.4523 & 0.600 & 37.55 & \textbf{17.89} & 51.48 \\ \hline
Single & 8 & 0.4474& 0.590 & 37.20& 19.67 & 53.78\\
& 16 & \textbf{0.4423} & \textbf{0.570}& \textbf{37.16}& 20.30& \textbf{54.20}\\
& 32 & 0.4758 & 0.600 & 33.41 & \textbf{18.55} & 46.60 \\
& 64 & 0.4804 & 0.590 & 34.08 & 19.79 & 43.75\\ \hline
\end{tabular}
}
\end{table}

\begin{table}[t!]
\centering
\caption{SELD performance of submitted models on dev-test data}
\label{tab:Submit_res}
% \resizebox{\columnwidth}{!}{%
\begin{tabular}{cccccc}
\hline
\textbf{Submission} &\multicolumn{1}{l}{\textbf{\begin{tabular}[c]{@{}l@{}}SELD\\ Score\end{tabular}}} & \multicolumn{1}{c}{\textbf{ER}} & \multicolumn{1}{c}{\textbf{F-score}} & \multicolumn{1}{c}{\textbf{LE}} & \multicolumn{1}{c}{\textbf{LR}} \\
 \hline
1 & 0.4023 & 0.495 & 42.68 & 16.70 & 55.19\\
2 & 0.4078 & 0.510 & 41.96 & 16.54 & 55.10\\
3 & 0.4149 & 0.515 &  41.24& 17.68&54.12 \\
\hline
\end{tabular}
% }
\end{table}

\section{Conclusion}
\label{sec:conc}

% In this technical report, DST attention is introduced for the SELD task, which proves to outperform the baseline model on all SELD metrics when applied with appropriate frequency pooling. Moreover, the incorporation of augmentation methods such as frameshift, time masking, channel swap, and moderate mix-up effectively addresses the issue of overfitting models to the training data. Furthermore, the utilization of external data synthesized using the balanced AudioSet dataset proves to be advantageous in accounting for diverse recording environments.

This technical report introduces DST attention for the SELD task, which demonstrates superior performance compared to the baseline model across all SELD metrics when combined with suitable frequency pooling. Additionally, the inclusion of augmentation methods such as frameshift, time masking, channel swap, and moderate mix-up effectively tackles the problem of overfitting to the training data. Furthermore, the integration of external data synthesized from the balanced AudioSet dataset proves to be beneficial in accounting for various recording environments.

\bibliographystyle{IEEEtran}
\bibliography{refs}

\end{sloppy}
\end{document}